\documentclass{article}

\usepackage{PRIMEarxiv}
\usepackage[colorlinks=true,linkcolor=blue,citecolor=blue,urlcolor=blue]{hyperref}
\usepackage[utf8]{inputenc} 
\usepackage[T1]{fontenc}    
\usepackage{url}            
\usepackage{booktabs}       
\usepackage{amsfonts}       
\usepackage{nicefrac}       
\usepackage{microtype}      
\usepackage{lipsum}
\usepackage{fancyhdr}       
\usepackage{graphicx}       
\graphicspath{{media/}}  
\usepackage{float}
\usepackage{longtable}

\title{Governed by Agents: A Survey on the Role of Agentic AI in Future Computing Environments
}

\author{
  Nauman Ali Murad \\
  Faculty of Computer Science \& Engineering \\
  GIK Institute, Topi, Pakistan \\
  \texttt{u2022479@giki.edu.pk}
  \And
  Safia Baloch \\
  Faculty of Computer Science \& Engineering \\
  GIK Institute, Topi, Pakistan \\
  \texttt{safia.baloch@giki.edu.pk}
}

\begin{document}
\maketitle

\begin{abstract}
The emergence of agentic Artificial Intelligence (AI), which can operate autonomously, demonstrate goal-directed behavior, and adaptively learn, indicates the onset of a massive change in today's computing infrastructure. This study investigates how agentic AI models' multiple characteristics may impact the architecture, governance, and operation under which computing environments function. Agentic AI has the potential to reduce reliance on extremely large (public) cloud environments due to resource efficiency, especially with processing and/or storage. The aforementioned characteristics provide us with an opportunity to canvas the likelihood of strategic migration in computing infrastructures away from massive public cloud services, towards more locally distributed architectures: edge computing and on-premises computing infrastructures. Many of these likely migrations will be spurred by factors like on-premises processing needs, diminished data consumption footprints, and cost savings. This study examines how a solution for implementing AI's autonomy could result in a re-architecture of the systems and model a departure from today's governance models to help us manage these increasingly autonomous agents, and an operational overhaul of processes over a very diverse computing systems landscape–that bring together computing via cloud, edge, and on-premises computing solutions. To enable us to explore these intertwined decisions, it will be fundamentally important to understand how to best position agentic AI, and to navigate the future state of computing infrastructures.

\end{abstract}

\keywords{Agentic AI \and Autonomous Systems \and Adaptive Learning \and AI Agents \and AI Governance \and Cloud Computing \and Edge Computing}

\section{Introduction}
Agentic AI systems are an advanced class of AI that are able to continuously learn, iteratively plan, and autonomously decompose tasks. While traditional AI chatbots are limited to moving quickly and automatically from task to task based on a user prompt, agentic AI goes further than just responding to prompts with answers—it also entails advanced reasoning processes and can independently execute steps in multi-step problems \cite{borghoff2025}. Agentic AI systems utilize cycles of perceiving, reasoning, acting, and learning in order to improve agent performance over time. This kind of system towards its own decision-making processes promotes the efficiency of processing and storage utility, as compared to traditional AI systems that typically think through brute-force computations \cite{openaiGovAI}. Currently, the potential for agentic AI crosses simple ideological significance of advancing technology and establishing agentic AI as a necessary step toward reliable human-oversight minimizing systems that could be deployed effectively and speculatively into the world of artificial intelligence \cite{researchgateAgenticAI}.

At the lowest levels, cloud computing offers the basic building blocks for AI deployments. From Infrastructure as a Service (IaaS) to Platform as a Service (PaaS) to Software as a Service (SaaS), the foundational layers have historically provided the flexible, scalable infrastructure necessary for LoRA deployments and in some cases, the supporting technology for LLMs. This is where agentic AI may shift an equilibrium in computing \cite{chen2024}. More and more, enterprises are reconsidering where to position agentic AI systems - on-premise, edge, specialized cloud vs. public cloud \cite{prangon2024}. This consideration arises from the potential of agentic AI to yield more value and cost via architectures which leverage fewer investments in high-end cloud resources and incorporate some localized processing capabilities \cite{sivakumar2024}. The main challenge is the availability of high-end processing and other cloud resource options offered by public cloud providers (production locked into massive GPU systems and storage systems, etc.) may be less desirable for agentic agent-based systems interested in making more efficient use of resources \cite{shetty2024}.

The rise of more autonomous and intelligent AI models currently raises many important questions about their effects on computing systems, from the cloud infrastructure to how we actually manage these systems \cite{khan2024}. These intelligent systems might eventually end up reshaping the way computing resources are handled, potentially leading to less dependence on public cloud services while also creating much more need for autonomous and fully secure deployment options across many different environments. Through some careful analysis, this research clearly examines how these AI systems are currently integrating with and transforming modern computing infrastructure, with particular focus on both the orchestration and security challenges, as well as the governance issues that come with decentralized AI control \cite{rehan2024}.

\section{Literature Review}
As agentic AI continues to evolve, it challenges long-held assumptions about how intelligent systems are built, deployed, and managed. To explore its broader implications, especially in the context of computing infrastructure, I’ve focused on literature that intersects AI autonomy, system architecture, and deployment models—each directly influencing how we might reimagine modern computing environments in light of agentic behavior.

Many studies clearly show that the traditional centralized cloud models just aren't enough when it comes to hosting autonomous agents. The work that \cite{chen2024} and \cite{prangon2024} have done points to both hybrid and edge computing as essential elements that help to reduce latency and enable more local decision-making. These architectural approaches end up delivering the contextual awareness and flexibility that agents need to fully operate in real-time.

The shift toward serverless and container-based environments is currently supporting this architectural change, as it provides much of the scalability and modularity that autonomous behaviors require \cite{samdani2022, shetty2024}. Additionally, these frameworks naturally complement the way that agentic AI is decentralized, which helps to facilitate quick adaptation and more streamlined orchestration

Managing autonomous systems presents some distinct challenges when it comes to governance. This is because agent behavior can be unpredictable and might be susceptible to many adversarial attacks, which means we need to have strong monitoring and accountability systems \cite{khan2024, javadi2020}. Moreover, these issues become much more complex in distributed environments where many users share resources and it becomes difficult to determine who is responsible for the outcomes.

The operational aspects of agentic systems, particularly their need for continuous learning, are currently calling for more innovative orchestration approaches. Much of the current research is looking into predictive AIOps and adaptive provisioning to help maintain consistent performance and to make sure that systems still remain responsive to changing contexts \cite{sivakumar2024, rehan2024}.
\vspace{-3mm} 

\section{Architectural Paradigms for Agentic AI Deployment} \label{sec:headings}

Agentic AI systems need many specific architectural patterns to fully function on their own across different environments like the cloud, edge, and on-premise settings. These essential patterns help to support much more complex workflows and eventually enable the seamless integration between various types of computing platforms. \cite{borghoff2025}.

\subsection{Foundational Agentic Design Patterns} 
The most important architectural patterns for agentic AI currently include the Reflection Pattern, Tool Use Pattern, Planning Pattern, and also the Multi-Agent Collaboration (MAC) Pattern. Each of these patterns clearly serves its own distinct purpose, ranging from enabling self-evaluation to helping facilitate collaborative problem-solving between many AI agents \cite{borghoff2025} As shown in Table~\ref{tab:my_label}. Moreover, these fundamental building blocks allow agentic AI systems to operate independently and eventually optimize their processes with just minimal human involvement \cite{chen2024}. 

\subsection{Architectural Models and Resource Implications} The careful selection of AI architecture still remains a very critical consideration, whether it might be vertical, horizontal, or some kind of hybrid approach, particularly when dealing with many different deployment scenarios \cite{prangon2024}. Although the final choice largely depends on both specific system requirements and various complexity levels, the inherent efficiency of agentic AI actually makes it much more suitable for smaller, localized systems rather than more extensive cloud infrastructure  \cite{samdani2022}.

\begin{figure}
    \centering
    \includegraphics[width=0.8\linewidth]{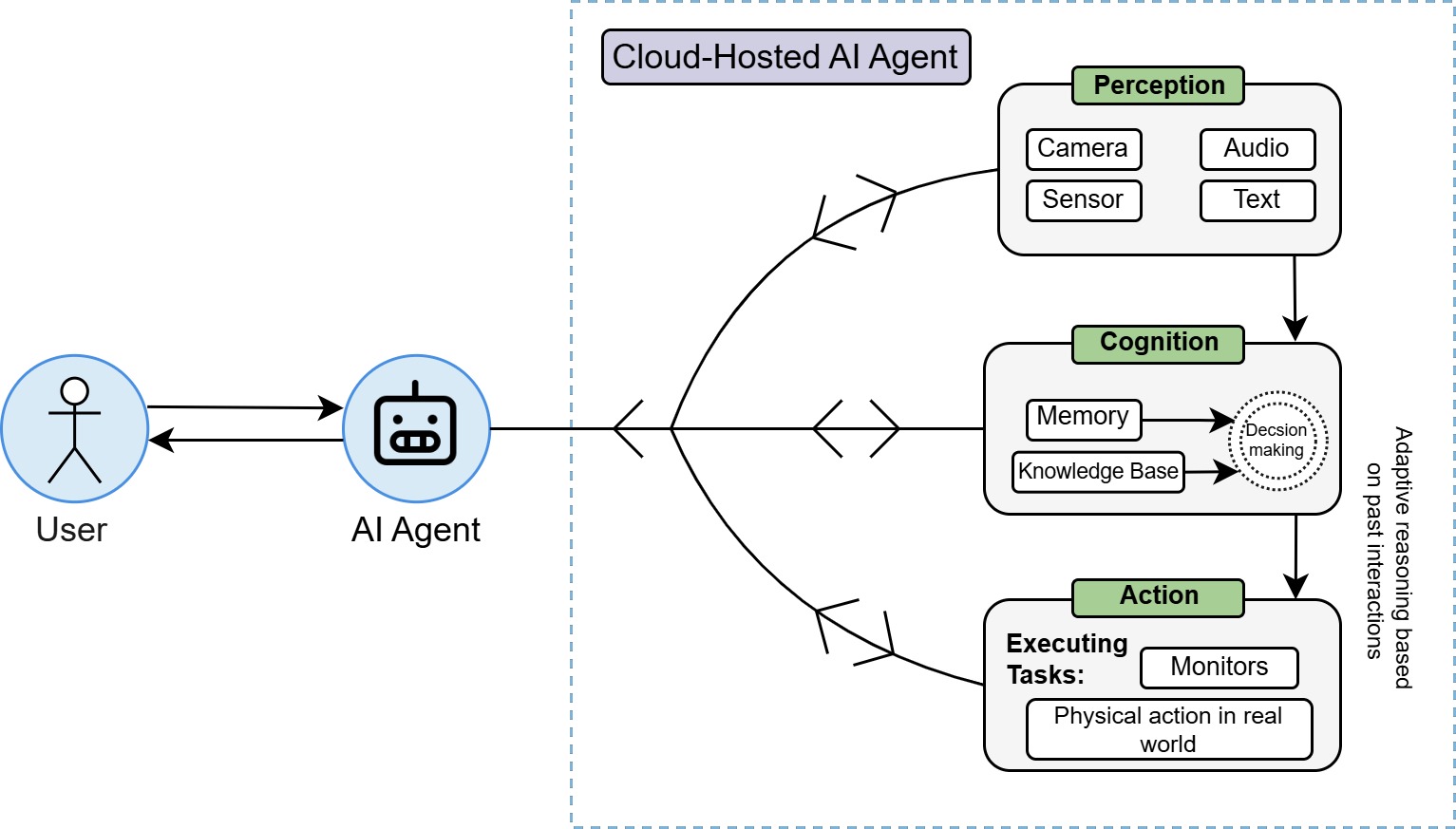}
    \caption{Architectural overview of an intelligent AI Agent system. The system has three core modules: Perception, Cognition, and Action, powered by cloud infrastructure.}
    \label{fig:enter-label}
\end{figure}
\vspace{-3mm} 

\subsection{Adapting Architectures for Edge and On-Premise Environments}

The implementation of systems that can act on their own, which actually have more localized processing abilities and need much less data, makes it possible to effectively use them in places where there might be limited resources, such as edge devices and some on-premise infrastructure. Unlike the more traditional way of just running AI workloads on flexible and highly scalable public cloud platforms, these kinds of systems can eventually operate with much more independence \cite{rehan2024}. Additionally, when teams are currently designing these types of architectures, it's clearly essential to fully consider the potential gaps that might occur in cloud connectivity and to make sure that each agent can still work autonomously by using the local computing resources for many extended periods of time.
\vspace{-2.5mm} 

\begin{table}[htbp]
    \centering
    \renewcommand{\arraystretch}{1.5}  
    \begin{tabular}{|c|p{3cm}|p{3.5cm}|p{3.5cm}|}
        \hline
        \textbf{Pattern Name} & \textbf{Description} & \textbf{Key Features} & \textbf{Use Cases} \\
        \hline
        Reflection Pattern & Self-evaluation and refinement of outputs & Iterative improvement, error correction, ambiguity clarification & Content creation, problem-solving, code generation \\
        \hline
        Tool Use Pattern & Interaction with external functions and APIs & Access to real-time data, extended functionalities, system integration & Tasks requiring external information or actions \\
        \hline
        Planning Pattern & Breaking down complex tasks into smaller steps & Task decomposition, strategic roadmaps, adaptive execution & Complex, multi-step projects, coding, logistics \\
        \hline
        Multi-Agent Collaboration (MAC) & Coordination among specialized agents & Diverse expertise, parallel processing, collaborative problem-solving & Complex, multifaceted tasks, scenarios requiring multiple viewpoints \\
        \hline
    \end{tabular}
    \newline
    \caption{Comparison of Architectural Patterns for Agentic AI}
    \label{tab:my_label}
\end{table}

\section{Frameworks and Platforms Enabling Agentic AI} 
Major cloud service providers, including Amazon Web Services (AWS), Microsoft Azure, and Google Cloud Platform (GCP), are actively developing and offering a range of tools and services to facilitate the building and deployment of agentic AI applications, while also acknowledging the potential for edge and hybrid deployments \cite{googleCloudAgentic}. Although these platforms clearly provide some robust infrastructure and many useful tools, the highly efficient nature of these AI agents means they might also work just as well on other types of platforms and systems \cite{prangon2024}.

\begin{table} [H]
    \centering
    \renewcommand{\arraystretch}{1.5}
    \begin{tabular}{|c|p{3.5cm}|p{3.8cm}|p{3.5cm}|}
        \hline
        \textbf{Provider/Framework} & \textbf{Platforms/Services} & \textbf{Description} & \textbf{Key Features} \\
        \hline
        AWS & Amazon Bedrock Agents, Amazon Q, Multi-Agent Collaboration, Model Context Protocol & Managed services and protocols for building, deploying, and managing autonomous and collaborative AI agents. & Easy setup, memory retention, built-in security, multi-agent orchestration, seamless integration with external data and tools. \\
        \hline
        Azure & Azure AI Foundry, Azure AI Agent Service, Semantic Kernel, AutoGen & Unified platform and frameworks for designing, customizing, and managing AI apps and agents, simplifying multi-agent orchestration. & Fully managed agent service, simplified orchestration, pre-built connectors, tools for building scalable and distributed agent networks. \\
        \hline
        GCP & Google Agentspace, Vertex AI Agent Builder, Agent2Agent (A2A) protocol & Platforms and protocols for building, deploying, and enabling interoperability between AI agents, focusing on enterprise data integration. & Unified platform for AI development, low-code agent building, secure communication between agents, access to enterprise data across various platforms. \\
        \hline
        Open Source Frameworks & LangChain, LangGraph, CrewAI & Versatile frameworks for building custom auto systems, managing workflows, and enabling multi-agent collaboration. & Modular design, stateful multi-actor systems, role-playing AI agents, extensive tool integration. \\
        \hline
        Specialized Vendors & Agentic AI, UiPath, Relevance AI, Salesforce Agentforce & Platforms offering AI agents as a service for various use cases with no-code/low-code interfaces. & Pre-built agents, easy deployment, integration with business systems \\
        \hline
    \end{tabular}
    \newline
    \caption{Comparison of Agentic AI Platforms and Frameworks}
    \label{tab:my_label}
\end{table}
\vspace{-8mm} 

\subsection{Cloud and Specialized Platforms for Agentic AI}
The major cloud providers, like AWS, Azure, and GCP, are currently moving much more toward adopting agentic AI through some of their platforms, which actually include services like Amazon Bedrock Agents, Azure AI Agent Service, and Vertex AI Agent Builder, because they want to make it possible for developers to create autonomous agents \cite{googleCloudAgentic}. Additionally, many frameworks, such as LangChain, LangGraph, and CrewAI, are now making it much easier to fully deploy and implement more streamlined and adaptable solutions on both affordable and local systems, which is clearly showing how the industry is eventually transitioning to more efficient and distributed agentic structures \cite{analyticsVidhyaFrameworks2025}.

\section{Agentic AI as a Service and the Shifting Cloud Landscape}

Cloud computing and agentic AI have actually become much more connected through what is currently known as Agentic AI as a Service, which essentially delivers AI capabilities by using cloud infrastructure. Many major providers, like AWS, Azure, and GCP, are still expanding their AI agent platforms to help boost both productivity and efficiency while also reducing some costs \cite{borghoff2025}. These services basically enable many organizations to build, deploy, and fully manage their AI agents through various cloud-based solutions \cite{googleCloudAgentic}.

The highly efficient nature of agentic AI is currently reshaping many traditional cloud computing models. Many organizations are now reconsidering their dependence on some major cloud providers that mostly just profit from selling extensive computing resources. Because agentic AI's design clearly emphasizes more local processing and much smaller data requirements, it might end up reducing the need for high-powered cloud infrastructure.

This transformation clearly opens up many possibilities for different types of computing approaches. Some companies might eventually shift toward on-premise systems, edge computing, and more specialized providers that are better suited for agentic AI workloads \cite{rehan2024}. While public clouds will still maintain their relevance for certain applications and hybrid solutions, the industry might see much more adoption of both private clouds and specialized AI platforms. Many managed service providers and hardware vendors are currently positioned to offer platforms that are specifically designed for agentic AI deployments, which basically reflects a broader shift in how cloud spending patterns are changing.

\section{Impact of Agentic AI on Computing Operations}

The way that AI agents are currently transforming how computing actually works is making it much easier to handle many different resources, like data and pricing. These systems are also moving away from the traditional cloud infrastructure to eventually create more flexible and efficient solutions.

These autonomous systems can now fully adjust the computing power whenever needed and actually fix many problems on their own, which makes operations much smoother across various computing environments. Additionally, they just don't need to rely as heavily on large cloud systems anymore, which clearly helps to reduce both costs and complexity.

When it comes to the management of data, AI agents are currently handling everything from processing to organizing with much more efficiency. They also tend to work particularly well with some smaller, more focused sets of data, which essentially means that less storage space might be needed and operations can end up becoming more streamlined.

This significant shift toward autonomous systems is currently creating many new ways to charge for services. Instead of using the traditional licensing model, businesses can now choose to pay based on either actual usage or specific tasks that are completed, which makes the costs much more directly related to the value they eventually receive from these AI systems.

\section{Future Directions and Outlook}

The rise of AI is currently set to transform many industries over the next decade, particularly through the development of more advanced agentic AI systems that might actually end up becoming just as impactful as cloud computing has been. This transformation will likely lead to some major changes in both how and where many computations are actually performed. Additionally, new business approaches will eventually start to emerge, like result-based services, while many existing models, such as SaaS, might also need to evolve to fully embrace more value-driven pricing structures.

Looking ahead, the future clearly points to both hybrid AI systems and much more enhanced edge computing capabilities. Many companies will likely need to balance their AI operations across the cloud, edge, and local systems to more effectively manage things like costs, speed, data control, and security. Moreover, this evolution suggests a clear move toward a more distributed AI infrastructure where autonomous agents can work across different computing environments, which might eventually lead to less dependency on some of the major cloud providers \cite{chen2024}.

\section{Conclusion}

The current integration of agentic AI has actually ended up transforming how we might approach both cloud and alternative types of infrastructures. Currently, there is clearly a trend that has emerged toward AI systems that can just independently reason, plan, and also act, which has led to much more improved automation and efficiency across many computing environments. 

The actual implementation of agentic AI still brings many significant challenges that need much careful attention. Currently, some balance must be struck between the AI capabilities and human oversight through well-designed governance frameworks that have to address ethical concerns about autonomy, accountability, and also value alignment \cite{openaiGovAI}. Moreover, the security landscape now demands more sophisticated threat modeling, enhanced detection systems, and also advanced identity management to fully protect against many vulnerabilities across different deployment models \cite{cyberarkAIsecurity, csaThreatModel}.

Cloud providers must now prepare for the evolving compute resource demands and also some new monetization approaches while still considering the growing significance of both edge computing and hybrid AI deployments. Additionally, the development of adaptive regulatory frameworks is currently essential for responsible innovation in agentic AI. Further, technical strategies should emphasize more flexible, scalable, and secure architectures across cloud, edge, and also on-premise environments. Taking a much more proactive stance on these challenges might actually facilitate the successful integration of agentic AI throughout many distributed computing systems \cite{googleCloudAgentic}.


\begin{thebibliography}{99}

\bibitem{borghoff2025}
U. M. Borghoff, “Human-Artificial Interaction in the Age of Agentic AI: A System-Theoretical Approach,” \textit{arXiv preprint arXiv:2502.14000}, Feb. 2025. [Online]. Available: \url{https://www.researchgate.net/publication/389176484}.

\bibitem{openaiGovAI}
OpenAI, “Practices for Governing Agentic AI Systems,” 2025. [Online]. Available: \url{https://cdn.openai.com/papers/practices-for-governing-agentic-ai-systems.pdf}.

\bibitem{researchgateAgenticAI}
Various, “AGENTIC AI: A COMPREHENSIVE FRAMEWORK FOR AUTONOMOUS DECISION-MAKING SYSTEMS IN ARTIFICIAL INTELLIGENCE,” \textit{ResearchGate}, 2025. [Online]. Available: \url{https://www.researchgate.net/publication/388188752}.

\bibitem{chen2024}
D. Chen, et al., “Transforming the Hybrid Cloud for Emerging AI Workloads,” \textit{arXiv preprint arXiv:2411.13239}, Nov. 2024. [Online]. Available: \url{https://arxiv.org/abs/2411.13239}.

\bibitem{prangon2024}
M. Prangon and J. Wu, “AI and Computing Horizons: Cloud and Edge in the Modern Era,” \textit{Journal of Cloud Computing}, vol. 13, no. 4, pp. 44, 2024. [Online]. Available: \url{https://www.mdpi.com/2224-2708/13/4/44}.

\bibitem{sivakumar2024}
R. Sivakumar, “Predictive AIOps for Intelligent Infrastructure Management,” \textit{International Journal of AI Operations}, 2024.

\bibitem{shetty2024}
A. Shetty, et al., “Lightweight Containers and Microservices for AI Workflows,” \textit{IEEE Transactions on Cloud Computing}, 2024.

\bibitem{rehan2024}
A. Rehan, “Adaptive Resource Provisioning in Edge-AI Environments,” \textit{Journal of Systems Architecture}, 2024.

\bibitem{khan2024}
A. Khan, et al., “Governance Models for Autonomous Multi-Agent Systems,” \textit{IEEE Access}, vol. 12, pp. 12345–12358, 2024.

\bibitem{samdani2022}
M. Samdani, et al., “Serverless Computing for Autonomous Systems: A Scalable Execution Model,” \textit{Proceedings of the 2022 International Conference on Cloud Engineering}, 2022.

\bibitem{javadi2020}
B. Javadi, et al., “Security and Accountability in Distributed AI Systems,” \textit{Future Generation Computer Systems}, vol. 108, pp. 823–835, 2020.

\bibitem{shetty2024}
A. Shetty, et al., “Lightweight Containers and Microservices for AI Workflows,” \textit{IEEE Transactions on Cloud Computing}, 2024.

\bibitem{cyberarkAIsecurity}
CyberArk, “The Agentic AI Revolution: 5 Unexpected Security Challenges,” 2025. [Online]. Available: \url{https://www.cyberark.com/resources/blog/the-agentic-ai-revolution-5-unexpected-security-challenges}.

\bibitem{csaThreatModel}
Cloud Security Alliance (CSA), “Agentic AI Threat Modeling Framework: MAESTRO | CSA,” 2025. [Online]. Available: \url{https://cloudsecurityalliance.org/blog/2025/02/06/agentic-ai-threat-modeling-framework-maestro}.

\bibitem{hiddenlayerSecuringAI}
HiddenLayer, “Securing Agentic AI,” 2025. [Online]. Available: \url{https://hiddenlayer.com/innovation-hub/securing-agentic-ai/}.

\bibitem{weaviateWorkflows}
Weaviate, “What Are Agentic Workflows? Patterns, Use Cases, Examples, and More,” 2025. [Online]. Available: \url{https://weaviate.io/blog/what-are-agentic-workflows}.

\bibitem{googleCloudAgentic}
Google Cloud, “Building the industry's best agentic AI ecosystem with partners | Google Cloud Blog,” 2025. [Online]. Available: \url{https://cloud.google.com/blog/topics/partners/best-agentic-ecosystem-helping-partners-build-ai-agents-next25}.

\bibitem{azureAIFoundry}
Azure, “New capabilities in Azure AI Foundry to build advanced agentic...,” 2025. [Online]. Available: \url{https://azure.microsoft.com/en-us/blog/new-capabilities-in-azure-ai-foundry-to-build-advanced-agentic-applications/}.

\bibitem{ibmWhatAreAIAgents}
IBM, “What Are AI Agents?,” 2025. [Online]. Available: \url{https://www.ibm.com/think/topics/ai-agents}.

\bibitem{analyticsVidhyaFrameworks2025}
Analytics Vidhya, “Top 7 Frameworks for Building AI Agents in 2025,” 2025. [Online]. Available: \url{https://www.analyticsvidhya.com/blog/2024/07/ai-agent-frameworks/}.

\end{thebibliography}
\end{document}